\documentclass[12pt]{article}
\input{epsf.tex}
\begin{document}
\begin{center}
{\bf LEVEL DENSITY AND RADIATIVE STRENGTH FUNCTIONS OF THE $^{237}$U NUCLEUS
FROM THE $(\overline{n},\gamma)$ REACTION
}\\
\end{center}\begin{center}{\bf V.A. Khitrov$^1$, 
A.M. Sukhovoj$^1$, V.M. Maslov$^2$}\\
\end{center}\begin{center}
{\it $^1$Joint Institute for Nuclear Research, Dubna, Russia\\

$^2$Joint Institute for Nuclear and  Energy Research, Minsk-Sosny, Belarus}\\
\end{center}

\begin{abstract}

       The independent analysis of the published data on the intensities of the
primary $\gamma$-quanta following resonance neutron capture in $^{236}U$ has been
performed. Distribution of these intensities about the mean value was approximated
in different energy intervals of the primary gamma-transitions and neutrons.
Extrapolation of the obtained functions to the zero registration threshold
of the primary gamma-transition intensity allowed us to estimate (independently on the
other experimental methods) expected level number of both parities for spin
values $J$=1/2, 3/2 and sum of radiative widths for both electric and magnetic
dipole gamma-transitions to levels with excitation energy up to $\approx 2.3$
MeV.
       Level densities and sums of radiative strength functions determined in
this way confirm characteristic behavior of analogous data derived from
intensities of the two-step cascades following thermal neutron radiactive
capture in nuclei from the mass-region $40 \leq A \leq 200$.
Besides, this permits one to estimate sign and magnitude of systematic
uncertainties for their model predicted values, at least, below one half of
the neutron binding energy. Comparison with the model notions of level density
testifies to super-liquid phase of this nucleus for the main part of excited
levels, at least, below 2.3 MeV.

\end{abstract}
\section{Introduction}\hspace*{16pt}

Precise models of level density $\rho$ and radiative strength functions
$k=\Gamma/(E_\gamma^3 D_\lambda A^{2/3})$
of the populated them primary dipole gamma-transitions at the neutron radiative
capture are necessary for estimation of experimental neutron cross-sections and
their calculation at lack of experimental data.
First of all, it is necessary for actinides.

Available \cite{RIPL} models, most probably, do not satisfy modern requirements
to these data \cite{Eval}.  Their insufficient quality is conditioned by clearly
small volume of reliable experimental data on level density and emission
probability, for example, of gamma-quanta. This results from significant errors
of experiments performed up to now. Underestimation  of uncertainties is largely
caused by the use of obsolete notions of a nucleus for analysis of experiment.

An increase of reliability level of the experimental data on $\rho$ and $k$
demands one to develop new methods for their determination at obliged
minimization of number of any assumptions and hypotheses of mechanism of used
nuclear reaction.

This condition is in high extent satisfied in analysis of the two-step
gamma-cascade intensities following thermal neutron capture \cite{Meth1}.
By means of this analysis it was observed for the first time the clearly
expressed step-wise structure with a width of $\sim 2$ MeV in the level density
$\rho$ below $\approx 0.5B_n$ in a large set of nuclei.
Further development of this method \cite{PEPAN-2005} brought to an additional
determination of unknown fact of strong dependence of partial widths
$\Gamma_{if}$ of not only primary but also following gamma-transitions on the
energy of the excited levels $f$ in the region of the structure mentioned above. 

Intensities $I_{\gamma\gamma}$ were measured up to now only for the case of
thermal neutron capture. Accumulated experimental data allowed one to determine
the $\rho$ and $k$ values for 42 nuclei \cite{Meth1}
(in the framework of standard hypothesis \cite{Axel, Brink} on independence of
radiative strength functions on nuclear excitation energy).
For two tens of them these data were obtained at relatively realistic
accounting for function $k(E_{\gamma}, E_{ex})$, experimentally estimated
below excitation energy $E_{ex} \sim 3-5$ MeV \cite{PEPAN-2005}.

It should be noted that the observed in \cite{PEPAN-2005} considerable
dtviation of relation
\\$R=k(E_{\gamma}, E_{ex})/k(E_{\gamma},E_{ex}= B_{n})=1$
indirectly and relatively weekly influences  the determined parameters of the
cascade gamma-decay.
This conclusion is true for the analysis of the intensities of the two-step
gamma-cascades to the final levels with excitation energy $E_{ex}<1$ MeV.
In this case, the maximum change in the calculated total radiative width of
cascade intermediate level (with accounting for function
$k(E_{\gamma}, E_{\rm ex})$ derived from experiment) brings to decrease in
the obtained according to \cite{PEPAN-2005} value of $\rho$ in region of
step-wise structure by a factor not more than 2 as compared with \cite{Meth1}.
Therefore, method \cite{PEPAN-2005} provides the $\rho$ and $k$ values with
the least at present systematic errors \cite{TSC-err}.

Nevertheless, obviously observed dependence of cascade intensity on the
structure of all three levels included initial compound-state \cite{Appr-k}
stipulates for necessity to get new experimental data on $\rho$ and $k$ from the
other experiments devoted to investigation of gamma-decay.
In addition, the methods \cite{Meth1,PEPAN-2005} belong to class of reverse
problems (determination of unknown parameters of functions measured in
experiment) and, that is why, they require maximum possible verification and
revealing all sources of systematic errors.
In practice, this condition stipulates for necessity to get additional set of
data on $\rho$
and $k$ from the maximum number of independent experiments. It is necessary also
to solve the problem of estimating possible dependence of found according to
\cite{Meth1,PEPAN-2005} gamma-decay parameters of high-lying levels on energy
of neutron resonances $\lambda$ and probable influence of their structures
on the process under study.
On the whole, analysis of the tendencies in determining $\rho$ and $k$ from
solution of reverse tasks points to necessity in further investigations aimed
to: 

\begin{itemize}

\item
estimation of adequacy of model notions of gamma-decay process to experiment 
and
\item
direct accounting for the coexistence and strong interaction between
Fermi- and Bose-systems in the radiative strength function models.
This must be done also in more details and more precisely in the level
density models for nuclei of any mass and type.
\end{itemize}

\section{Experimental data from the $(\overline{n},\gamma)$ reaction and
method of their analysis}

Both solution of the problem of correctness of model notions and necessity to
study influence of the structure of the initial compound-states on the
cascade gamma-decay process is up to now the primary aim of experiment.
At present, this can be done in analysis of experimental intensities of
the primary gamma-transitions following capture  of ``filtered" neutrons with
energy 2 and 24 keV (or by any averaging over resolved resonances).

The most complete set of these data among actinides is accumulated for
compound nucleus $^{237}$U.
Unfortunately, authors of corresponding experiments \cite{U237} used their
data practically only for determination of spin and parity of excited levels
on the base of notions of the limited ``statistics" theory of gamma-decay.
I.e., in the framework of the hypotheses:
       
\begin{itemize}
\item
on independence of $k(E1)$ and $k(M1)$ on structure of decaying ($\lambda$) and
excited ($f$) nuclear levels and 
\item
applicability of the Porter-Thomas distribution \cite{PT} for describing random
deviations of the gamma-transition partial widths in any interval of their
energy from mean values.
\end{itemize}

There are no experimental evidences of hypothesis  \cite{Axel, Brink, PT}
for the data like \cite{U237} (concrete nucleus, given set of gamma-transition
intensities). Therefore, the method of analysis must take into account
possibility of non-execution of assumptions mentioned above.

\subsection{Algorithm of analysis}
The analysis is based on the following statements:
\begin{itemize}

\item
the number of the primary dipole transitions observed in the
$(\overline{n},\gamma)$ reaction is less than or equal to the value 
$\rho \times \Delta E$ for any excitation energy interval $\Delta E$ and spin
window determined by the selection rule on multipolarity;

\item
the sum of widths of the observed transitions is less than or equal to the
summed width of all the possible primary  gamma-transitions;
\item
the likelihood function of approximation of  distribution of random intensity deviations from the mean value has the only maximum;

\item
the experimental values of $\rho \Delta E$ and sum $\Gamma_{\lambda,f}$ can be
determined with acceptable uncertainty from extrapolation by curve which
approximates distribution of the random gamma-transition intensities to zero
threshold of their registration;

\item
the averaging of random fluctuations of the primary gamma-transition widths over the initial compound states decreases dispersion of their distribution
independently on reliability of hypothesis \cite{PT}. I. e., any set
of gamma-transition intensities from the $(\overline{n},\gamma)$ reaction can
be approximately described by the $\chi^2$-distribution with unknown number of
degrees of freedom $\nu$. (In practice this distribution is close to normal one
with dispersion $\sigma^2 \approx 2/\nu$.)
\end{itemize}

This means that the number of the primary gamma-transitions from reaction 
$(\overline{n},\gamma)$ with intensity lying below detection
threshold of experiment \cite{U237} in any case considerably decreases as
compared with analogous data obtained for decay of the only compound-state.
In this case, determination of the primary gamma-transition intensity distribution parameters provides  maximum possible precision for
extrapolation of function of intensities random values distribution 
into the region below the detection threshold. 

Therefore, precision in estimating the number of unobserved gamma-transitions
and sum of their partial widths from the data \cite{U237} must be much higher
than it can be reached by analysis of intensities of primary gamma-transitions
following thermal neutron capture. Corresponding technique was developed
earlier and tested on large set  of the data on the two-step cascade intensities
in \cite{Appr-TSC}.

The problem of principle importance in the analysis of such kind is really unknown law of distribution of dispersion of random widths about mean value.
The fact of discrepancy between dispersion of random intensities of primary
gamma-transitions and expected for a given nucleus value of $\nu$ at the capture
of 2 keV neutrons was first pointed out in \cite{Yb172}.
But up to now attempts to solve this problem were undertaken up to now.

The Porter-Thomas distribution correctly describes distribution of the random
partial widths of the tested gamma-transitions only when their amplitudes have
normal distribution with zero mean value.
Therefore, they must be the sum of large number of items of different signs and
the same order of magnitude. This condition must be fulfilled if wave-functions
of the levels connected by gamma-transition contain a large amount of items
with different signs and comparable magnitudes.
Just these components are contained in matrix element for amplitude of
gamma-transition.

\subsection{Some aspects of modern theoretic ideas of gamma-transition
probability}\hspace*{16pt}

On the whole, existing theoretic developments, for example,
quasiparticle-phonon nuclear model (QPNM) \cite{PEPAN-1972,QPNM} call some
doubts about applicability of mentioned above primitive ideas of gamma-decay.
In particular, the regularities of fragmentation of the different complicated
states studied in the frameworks of QPNM \cite{MalSol} directly point to
presence of items with considerable component of wave functions in the
primary transition amplitudes. First of all, this concerns wave functions of
excited levels \cite{Even-Odd}, but it is not excluded that the wave-functions of decaying compound states (neutron resonances) also have large components
\cite{PEPAN-1972}. This directly results in potential possibility of rather
considerable violation of the Porter-Thomas distribution.
These violations can appear themselves in limited \cite{MalSol} energy intervals of final levels and change dispersion of real distribution as compared with
\cite{PT}. Correlation between absolute values of items in amplitude of any
gamma-transition and their signs for the data like \cite{U237}
are unknown. Therefore, the suggested below analysis of available experimental
data must take into account possibility of strong dependence of the primary
transition partial widths on structure and, correspondingly, energy of excited
levels and cover all spectrum of their random deviations from the mean value.
At least, it must guaranty obtaining the minimum possible estimation for $\rho$
and the maximum possible -- for $k$. Just this sign of deviation of their
experimental
values from the practically used \cite{RIPL} model ideas is provided by the
analysis \cite{Meth1,PEPAN-2005}.

According to theoretical notions of QPNM, the amplitude of gamma-transition  from
high-excited nuclear state (neutron resonance) is a sum of different in
structure elements \cite{PEPAN-1972}. Schematically \cite{Malov} they consist
from a number of items which correspond to the following components of
wave-functions of decaying and excited levels connected by the gamma-transition:

(a) n-quasiparticle,

(b) n-quasiparticle $\otimes$ phonon, 

(c) n-quasiparticle $\otimes$ two phonons and so on.

I. e., amplitude of given gamma-transition can be determined by some components
of physically different types. In common case they can have considerably different scale.
And concrete values are determined by degree of fragmentation of the nuclear states
enumerated above. The types of dominant components in the wave-functions of
final levels excited by primary gamma-transitions can be different in principle,
especially at low excitation energy of levels \cite{Even-Odd}.

The first part of amplitude (see, for example, \cite{Malov}) for many rather
high-lying levels is determined by a number of items of different sign and,
on the average, comparable magnitude. This qualitative explanation follows from calculation of
structures of low-lying levels of deformed nuclei performed by authors  
\cite{Even-Odd} and the most general principles of fragmentation of nuclear
states of complicated structure as increasing excitation energy \cite{MalSol}.

The following items in the gamma-transition amplitude (at sufficient energy of
excited level) account for contribution of those components which cause change
in wave functions by one phonon. I.e., it follows from main theses of QPNM,
for example, that the all multitude of the primary gamma-transition amplitudes
cannot be reduced to one limited case as it was suggested in \cite{PT}.

\subsection{Some  peculiarities of experimental data}
\hspace*{16pt}
Experimental investigations of various target-nuclei were performed in BNL.
But only even-odd target-nucleus $^{236}$U was chosen for the analysis
presented below. This choice is stipulated by maximum interval of the primary
gamma-transition energies listed in  \cite{U237}.

Even-odd compound-nucleus has an only possible spin at capture of s-neutrons
and two possible -- for p-neutrons. Analysis of intensities in the last case
requires one to introduce and then determine the number of parameters
corresponding to even-even nucleus. Even-odd compound nucleus at its excitation
by s-neutrons represents methodically  a particular case of the task considered in \cite{Yb174}.

The width FWHM=850 eV of filtered neutron beam with the energy of 2 keV in the
performed experiments was determined by interference minimum in the total
cross section of scandium. The average spacing between neuron resonances in
$^{236}$U equals 12 eV and this provides minimum number of the primary
gamma-transitions whose intensities are less than detection threshold.

If one does not account for:

(a) the change in neutron flux in the energy interval mentioned above;

(b) the possible strong correlation of partial radiative widths and 

(c) the presence of noticeable statistic errors in experimental data

then dispersion  $\sigma^2=2/\nu$ of their expected distribution can be not
less than $\sim 0.03$. In presence of absolute correlation between reduced
neutron width and partial radiative widths one can estimate maximum possible
dispersion from the folding of two $\chi^2$ dispersions by the value
$\sigma^2=8/\nu \geq 0.12$.

I.e, the main part of the primary gamma-transition intensities observed in
the nucleus under consideration must exceed experimental detection threshold.
Therefore, expected errors of extrapolation must be small enough.

It is assumed in analysis that all the distributions of the primary
gamma-transition intensities from reaction $(\overline{n},\gamma)$ have only
the following unknown parameters:

(a) the averaged reduced intensity $\langle I_\gamma^{max}/E^3_\gamma\rangle$ of
gamma-transitions populating levels $J=1/2,3/2$;

(b) the portion $B=\langle I_\gamma^{min}\rangle/\langle I_\gamma^{max}\rangle$ of reduced intensities of gamma-transitions to levels $J=5/2$ relatively  to that to levels $J=1/2,3/2$
(practically
- for intensities of primary transitions following capture of neutrons with
energy of 24 keV;

(c) the ratio $R_k=k(M1)/k(E1)$ which is independent on spin values of the levels
populated by the primary transitions;

(d) the expected and equal numbers $N_\gamma$ of gamma-transitions to levels
$J=1/2,3/2$ and $J=5/2$;

(e) as well as the dispersion $\sigma^2$ measured in units of degree of freedom
$\nu$.

Naturally, these parameters are to be determined independently for each energy
interval of the primary gamma-transitions.

The presence of statistical errors in determination of each experimental value
of $\langle I_\gamma/E^3_\gamma\rangle$ automatically increases experimental dispersion
$\sigma^2$ of distribution and decreases the $\nu$ value. It is assumed
that their relative systematic errors in each energy interval are practically
equal.

Of course, this notion assumes that the structures of initial compound-state
and a group of levels in rather narrow interval $\Delta E$ of excitation energy
connected by the primary gamma-transitions of the same type weekly influence
the mean reduced intensities $\langle I_\gamma/E^3_\gamma\rangle$ of these quanta.

Both performed in \cite{Appr-k} approximation and interpretation of
experimental data on $k(E1)+k(M1)$ and ideas of modern nuclear theories show
that this assumption can contain considerable uncertainty (especially for wide
energy intervals of the primary gamma-transitions under study).
But the maximum accuracy in determination of the most probable values of
$N_\gamma$, $B$, $R_k$, $\nu$ and  $\langle I_\gamma^{max}\rangle$ can be achieved,
in principle, by recurrent  optimization of the primary gamma-transition energy
intervals where are determined these parameters.

One more problem is due to small volume of the set and  difference in numbers
of electric and magnetic dipole gamma-transitions in concrete intervals
$\Delta E$. Therefore, it is necessary to introduce and fix in analysis some
assumption about number of levels of positive and negative parity in a given
energy interval of nuclear levels. Below is used the hypothesis of equality of
number of electric and magnetic gamma-transitions. In practice, this ratio
can be varied for any possible hypotheses of ratio between level densities
with different parity for any given excitation energy interval.
The problem of difference in level density of different parity disappears
for values of $R_k \approx 1$, the maximum error in determination of $N_\gamma$
in case $R_k \approx 0$ corresponds to the lowest intensity transitions
and insignificantly distorts  desired sum $\sum \langle I_\gamma/E^3_\gamma\rangle$.
In intermediate case, error of approximation will be stipulated, first of all,
by difference in level densities of positive and negative parity -- it will
decrease as increasing excitation energy (as it was on the whole predicted by
modern theoretical calculation  of this nucleus parameter \cite{PEPAN-1976}).

Approximation of the mixture of the different type random values
with respect to mean parameters
by any distribution cannot determine their belonging to
certain type without using additional information. But, accounting for the
known fact that the magnetic gamma-transitions to the lowest levels are by order
of magnitude weaker than electric transitions, one can extrapolate inequality
$R_k=k(M1)/k(E1)<1$ for the nuclei under study up to excitation energy where
$R_k=1$. There is not excluded that at higher energies of gamma-quanta
$k(M1)/k(E1)>1$.

Strength function of p-neurons $S_1=2.3(6)$ in isotope $^{236}$U noticeably
exceeds strength function of s-neutrons $S_0=1.0(1)$ \cite{BNL-325}.
Authors of \cite{U239} estimated that in this case the portion of captures of
p-neutrons with energy of 2 keV equals approximately 15\%.
If one does not account for possible irradiation of small group of the primary
dipole gamma-transitions following capture of p-neutrons and terminating at
the levels with spin values $J=5/2$, then the presence of this capture appears itself, most probably, as change in the $R_k$ values for different energies of excited levels and corresponding increase of $\nu$.
Therefore, the presence of small number of the 2 keV p-neutron captures must not noticeably influence accuracy in determination of the
expected values of $N_\gamma$ and sum  $k(E1)+k(M1)$.

There is no problem of p-neutron capture for the data on the intensities of primary gamma-transitions for resolved resonances.
Absolute normalization of partial widths for each resonance performed in \cite{U237}
allows one to provide for the most effective averaging of them.
Unfortunately, arithmetic mean value of $\Gamma_{if}$ inevitably shifts to
lower values by different for each gamma-transition quantity
$M \times \Delta\Gamma_{if}$.
Partial width $\Delta\Gamma_{if}$ of each from  $M$ gamma-transition with
intensity lying below registration threshold in given resonance varies  from
zero to some maximum magnitude. That is why, averaging over resonances
additionally has this unknown specific error.

Approximation of distribution of random values $I_\gamma/E^3_\gamma$ was
performed by analogy with \cite{Appr-TSC} for cumulative sums in function
of increasing values of intensities.

\section{Results of analysis}\hspace*{16pt}

The examples of experimental distributions of cumulative sums of the primary
transition reduced intensities
$\sum I_\gamma/E^3_\gamma= F(\langle I_\gamma/E^3_\gamma\rangle, N_\gamma, \nu, R_k)$ were
calculated for different values of concrete parameters and given in \cite{Yb174}.

\begin{figure} [htbp]
\begin{center}
\leavevmode
\vspace{4.cm}
\epsfxsize=16cm
\epsfbox{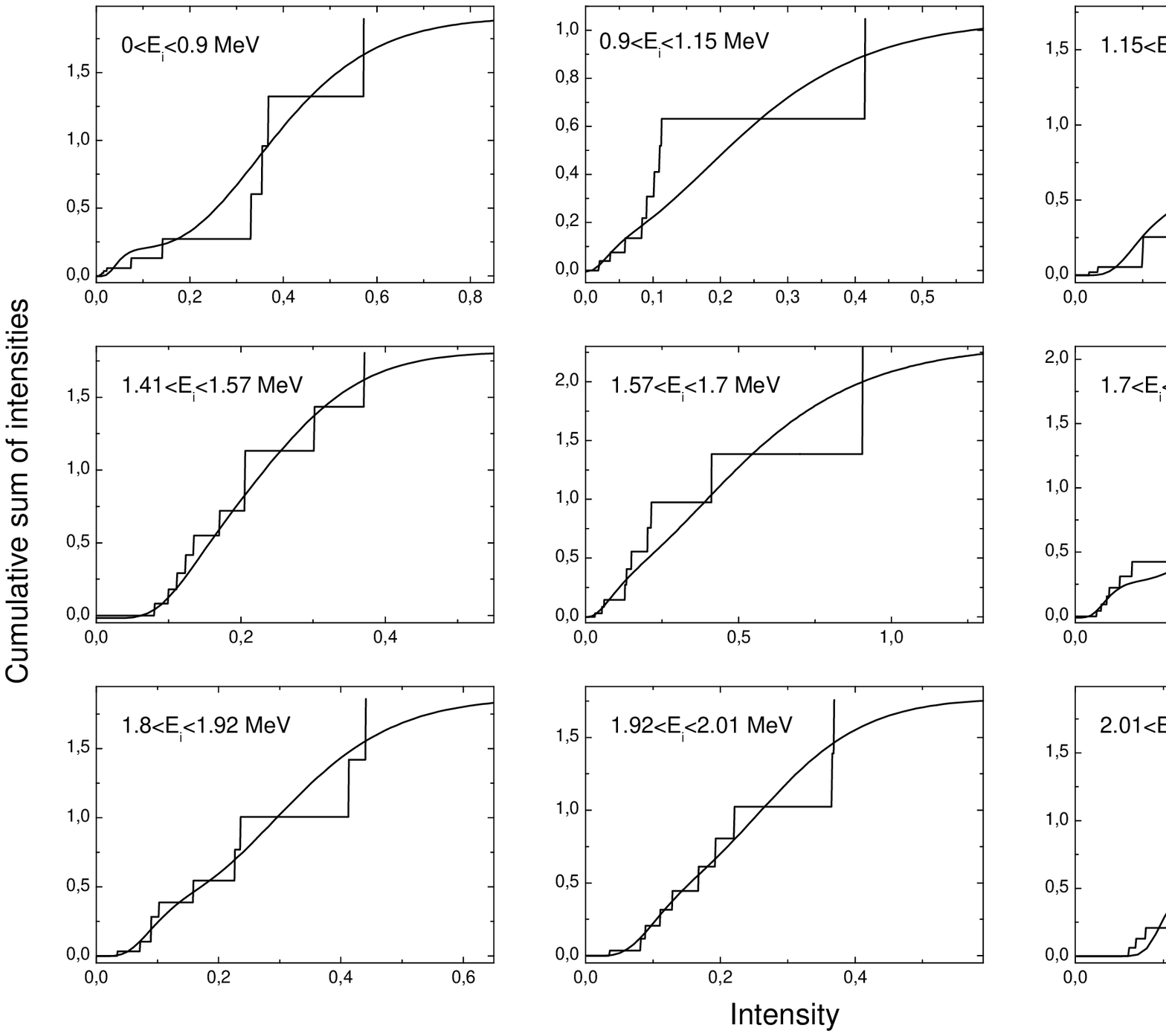}
\end{center}

\vspace{-5.5cm}

{\bf Fig.~1.} The histogram represents experimental cumulative sum of reduced
relative intensities $\langle I_\gamma\rangle/ E_\gamma^3$ for $^{237}$U.
Smooth curve shows the best approximation.
The intervals of excitation energy $E_i$ of final nuclear levels are shown
in figures. Experimental data for neutron energy $5\leq B_n\leq 125$ eV. 
\end{figure}

\begin{figure} [htbp]
\vspace{-2.5cm}
\begin{center}
\leavevmode
\epsfxsize=9cm
\epsfbox{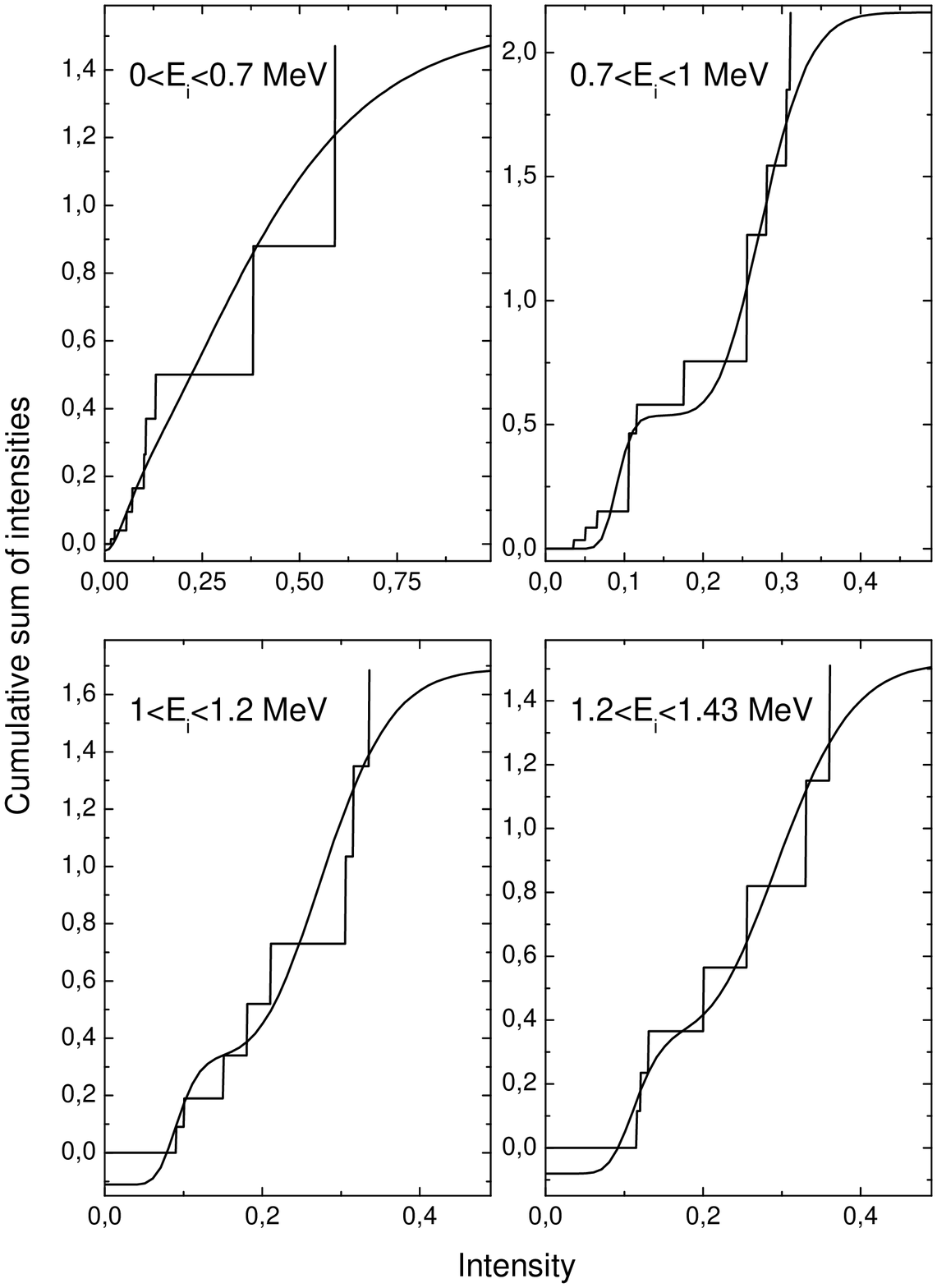}
\end{center}
\vspace{-1.cm}

{\bf Fig. 2.} The same, as in Fig. 1, for $E_n \approx 2$ keV.

\end{figure}
\begin{figure} [htbp]
\vspace{-2.5cm}
\begin{center}
\leavevmode
\epsfxsize=9cm
\epsfbox{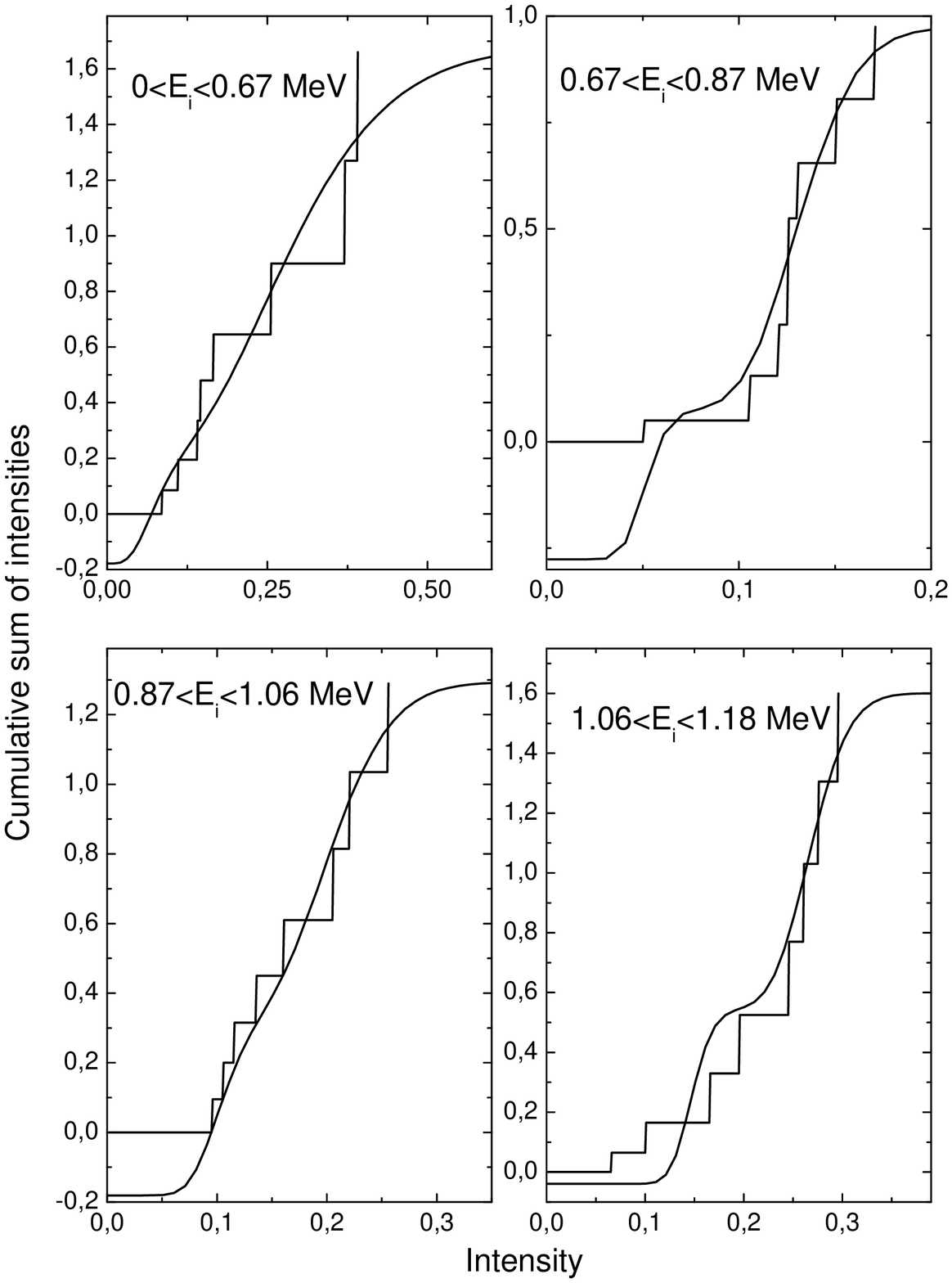}
\end{center}
\vspace{-1.cm}

{\bf Fig. 3} The same, as in Fig. 1, for $E_n \approx 24$ keV.
\end{figure}

This was done only under condition of correspondence  between experiment and
accepted hypotheses of the distribution form of the primary transition
random intensities. The presence of functional dependence of the primary
transition intensities on some ``hidden" parameter can result in maximal errors of approximated values of the most probable number of
gamma-transitions and dispersion of their deviation from mean value.
Modern nuclear theory does not consider this possibility; there are no
experimental data on existence of ``hidden" dependence, as well.
Therefore, it is not taken into account below.

Experimental cumulative sums of the $\langle I_\gamma/ E_\gamma^3\rangle$ relative values
are shown in figs. 1-3 together with their best approximation.
The data are presented so that the expected total intensity of
gamma-transitions lying below detection threshold corresponds to the most
probable value of cumulative sum for $\langle I_\gamma/ E_\gamma^3\rangle=0$.

At low energy $E_i$ of final levels, accuracy in determination pf parameters
of approximating curve must get worse due to inequality of level density with
different parity. Most probably, this increases error of extrapolation of
gamma-transition intensities to zero value. This can result in overestimating
of $N_\gamma$ values.

The best values of fitting parameters $\nu$ and $R_k$ are are given in figs. 4
and 5. Noticeable change in these parameters of approximation for $0.7<E_i<1.2$
MeV points to considerable change in structure of given even-odd isotope
in this excitation energy region.

\begin{figure} [htbp]
\vspace{4 cm}
\begin{center}
\leavevmode
\epsfxsize=16.5cm
\epsfbox{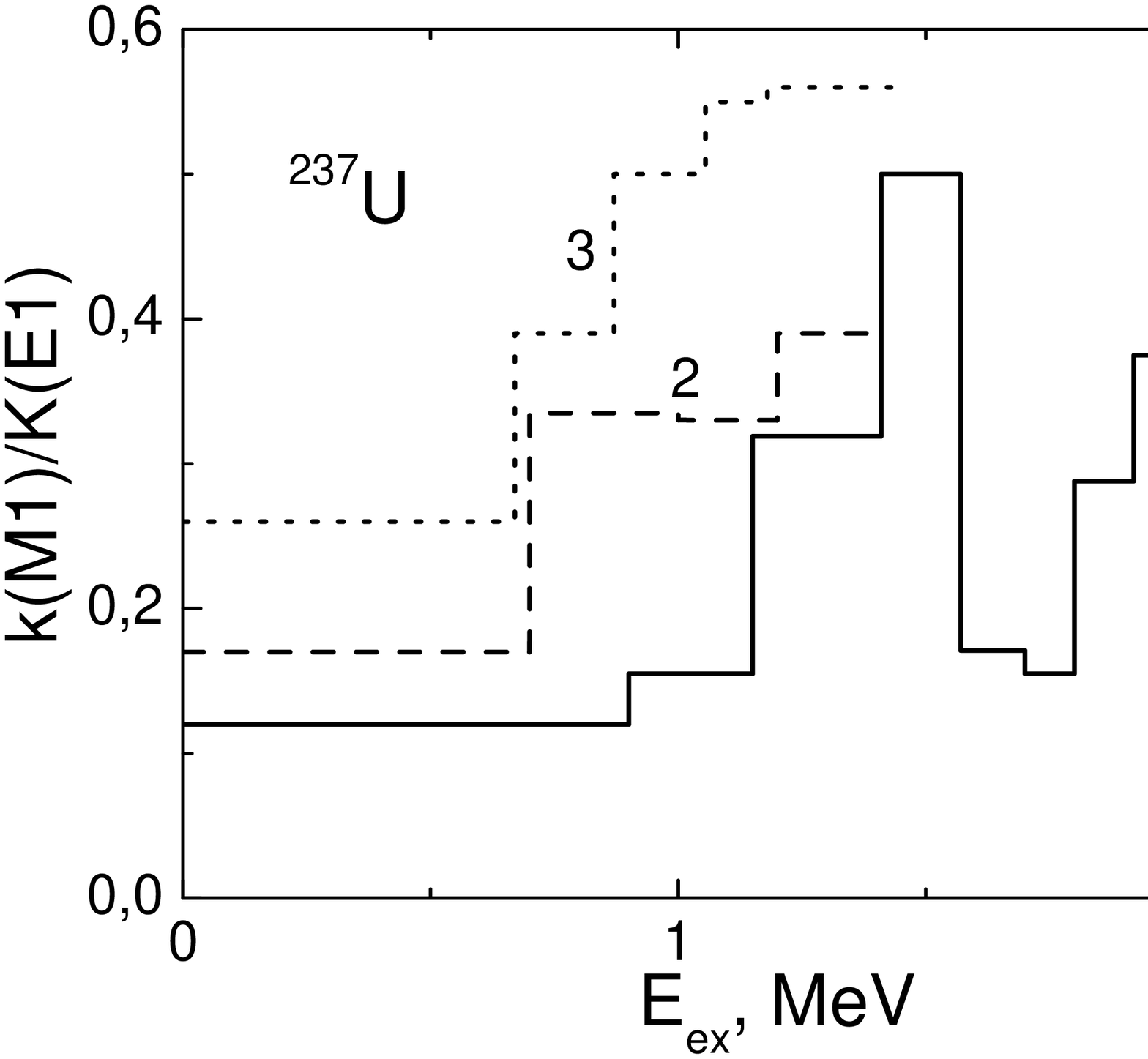}
\end{center}

\vspace{-5.5cm}

{\bf Fig.~4.} The histograms represent the best values of ratio
$k(M1)/k(E1)$ for different excitation energies of levels populated by
dipole gamma-transitions.
Line 1 represents data for $5<E_n<125$ eV, line 2 - for $E_n \approx 2$
keV and line 3  -- for $E_n \approx 24$ keV.

\end{figure}

\begin{figure} [htbp]
\vspace{3 cm}
\begin{center}
\leavevmode
\epsfxsize=16.5cm
\epsfbox{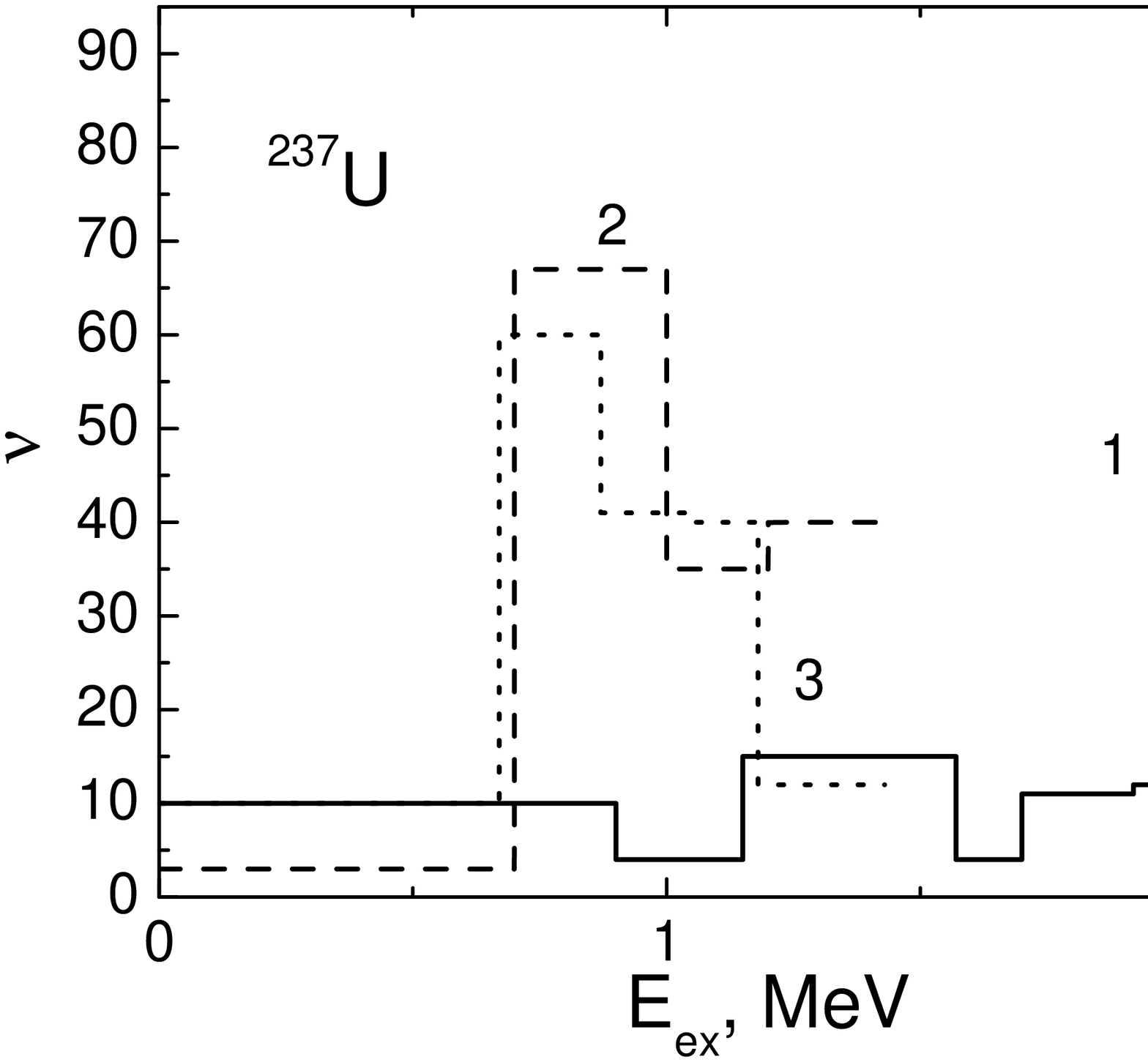}
\end{center}
\hspace{-0.8cm}
\vspace{-6.cm}

{\bf Fig.~5.} The same, as in Fig.4, for parameter $\nu$.
\end{figure}

The best values of level density $\rho = \sum_{J,\pi}N_\gamma/\Delta E$
and summed radiative strength functions
$\sum \langle I_\gamma\rangle/ (E_\gamma^3 N_\gamma)$ are given in figs. 6 and 7.
Normalization of intensities and strength functions in both \cite{Meth1} and
\cite{U237} was done to their absolute values.
Because gamma-transition intensities following capture of ``filtered" neutrons
are listed in \cite{PT} in relative units than corresponding strength functions 
Fig.7 are combined with ``resonance" values under assumption of their
approximate equality.

\begin{figure} [htbp]
\vspace{3 cm}
\begin{center}
\leavevmode
\epsfxsize=16.5cm
\epsfbox{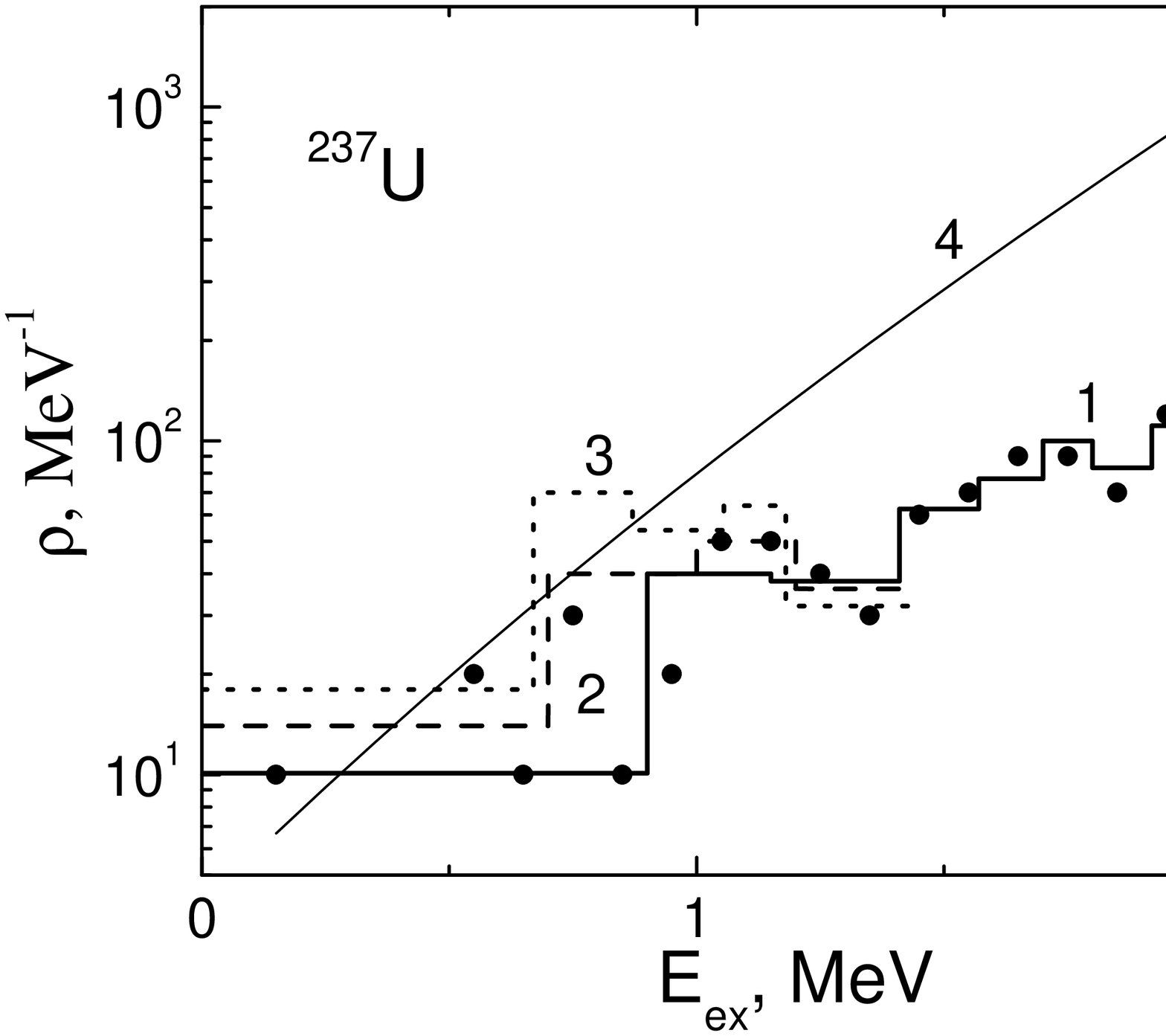}
\end{center}
\vspace{-5. cm}

{\bf Fig.~6.} The same, as in Fig. 4, for the density of excited levels.
Line 4 represents calculation within model \cite{Dilg} for spin values J=1/2 and 3/2.
Points correspond to number of levels observed in resolved resonances.

\end{figure}
\begin{figure} [htbp]
\vspace{3 cm}
\begin{center}
\leavevmode
\epsfxsize=16.5cm
\epsfbox{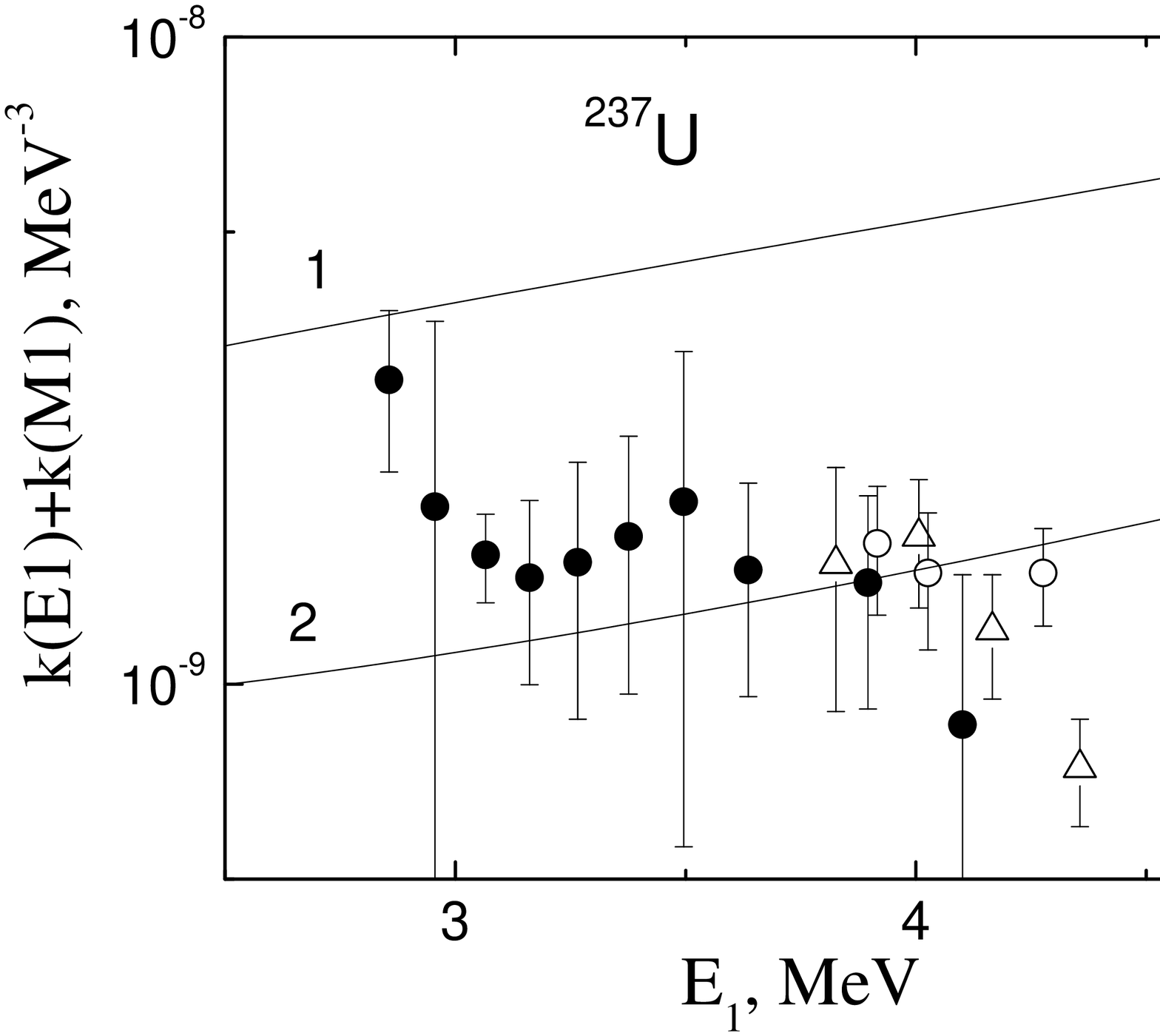}
\end{center}

\vspace{-5.cm}

{\bf Fig.~7.} The summed radiative strength functions derived from the
data on capture of neutrons with different energies $E_n$: black points --
$5 < E_n < 125$ eV, open points -- $E_n \approx 2$ keV,
triangles -- $E_n \approx 24$ keV.
Line 1 represents calculation within model \cite{Axel},
line 2 -- calculation within
model \cite{KMF} together with k(M1)= const.

\end{figure}

\subsection{Some sources of systematic errors}\hspace*{16pt}

Absolute minimum of $\chi^2$ for all used sets of intensities is achieved
practically for the only $N_\gamma$ value. Change in this parameter by $\pm 1$ brings to significant increase in $\chi^2$.
This allows one to neglect possibility of considerable (for example, 10-20\%) error of determined level density. 

Main problems in determination of nuclear parameters and their systematic
errors, most probably, are stipulated by:

(a) the use of assumptions on the distribution form of the random intensity
deviations from the unknown mean value and

(b) possible presence of significant errors in the set of the intensities under
analysis.

1. The Porter-Thomas distribution allows very considerable random deviations
of partial widths. But the measured gamma-transition intensities are always 
limited by finite value of the total radiative width of decaying state.
Therefore, the set of the best values of parameters depends on the total
width of region for approximation of cumulative sums.
Mainly this concerns the best value of parameter $\nu$.
In the performed analysis, intensities were normalized so that their maximum
value would not exceed 50-70\% of the approximation region width.

2. The main error of analysis can be related only to the ``loss" of
gamma-transitions whose intensities exceed threshold value and/or due to mistaken
identification of the secondary gamma-transitions as the primary one.
The probability of overlapping of two peaks corresponding to neighbouring
levels was estimated in \cite{Stel}. As it follows from the data presented
by authors, this effect
is rather small and, most probably, cannot explain considerable
(several times) discrepancy between level density determined by us and
predictions of the Fermi-gas level density model \cite{Dilg}.

Considerable uncertainty could be due to even and significant loss of 
observed peaks corresponding to intense primary transitions
(exceeding their detection threshold) owing to their grouping in multiplets
with rather narrow ($\sim 1-2$ keV) spacing between peaks.
But this possibility is not predicted by modern nuclear models.

3. In principle, there is possible the situation when gamma-transitions in
all or larger part of chosen intervals of primary transition energies
(with the width of some hundreds keV) have different mean values.
Moreover, probability of relatively low-intense transitions strongly but
smoothly increases as decreasing their intensities. Potentially, this effect can be due to by fragmentation mechanism of states over concrete neighbouring nuclear levels. 

Apparently, only such hypothesis can be alternative potential explanation of
``step-wise" structure in level density in analysis performed in
this work.
Application of this hypothesis to the level densities determined according
to \cite{Meth1,PEPAN-2005} requires that the main portion of levels 
with the same $J^{\pi}$ below
$\approx 0.5B_n$ would not be excited by the primary gamma-transitions.
Besides, some Cooper nucleon pairs would break simultaneously at low in
comparison with $B_n$ nuclear excitation energy.
We cannot suggest other possibility for precise reproduction 
of the two-step gamma-cascade intensities in calculation.

\subsection {Interpretation of the obtained results}\hspace{16pt}
\begin{figure} [htbp]
\vspace{4 cm}
\begin{center}
\leavevmode
\epsfxsize=16.5cm
\epsfbox{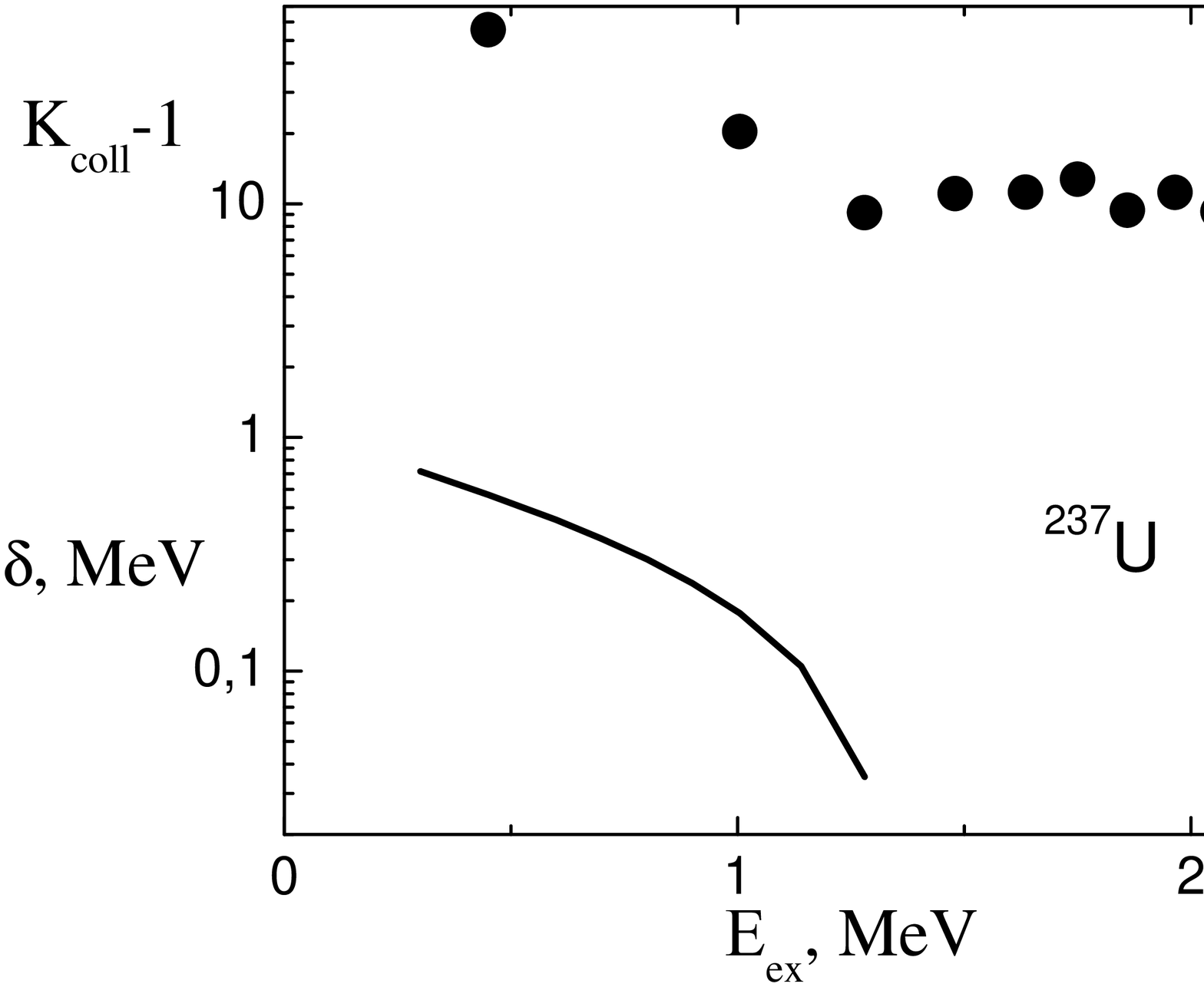}
\end{center}

\vspace{-5.5cm}

{\bf Fig.~8.} The points represent the coefficient of collective enhancement
of level density, the line shows the values of parameter $\delta_1$ used
in \cite{Meth1,PEPAN-2006} for calculating partial density of
three-quasiparticle levels.
\end{figure}

The physics important information on level structures below $\approx 0.5B_n$
can be extracted from the values of coefficient of collective enhancement of
level density:
\begin{equation} \rho(U,J,\pi)= \rho_{\rm qp}(U,J,\pi) K_{\rm coll}(U,J,\pi).
\end{equation}

According to modern notions, $K_{\rm coll}$ determines in deformed nucleus
\cite{RIPL} a degree of enhancement of level density of pure quasi-particle
excitations $\rho_{\rm qp}(U,J,\pi)$ due to its vibrations and rotation.
In very narrow spin window considered here one can accept in the first
approach that, to a precision of small constant coefficient, it equals
coefficient of vibration  increase in level density $K_{\rm vibr}$.
On the whole, the last is determined by change in entropy of a nucleus
$\delta S$ and redistribution of the nuclear excitation energy $\delta U$
between quasi-particles and phonons at nuclear temperature $T$:

\begin{equation}
K_{\rm vibr}=exp(\delta S-\delta U/T).
\end{equation}

The experimental data at hand on level density and existing model notions of
it do not allow one  unambiguous and reliable determination of the $K_{\rm vibr}$
value for arbitrary nuclear excitation energy $U$ even at zero systematic
error in determination of function $\rho(U,J,\pi)$.

Now  is no possibility for ambiguous  experimental determination \cite{Prep196}
of breaking threshold $E_N$ of the first, secondary and following Cooper pairs,
value and form of correlation functions $\delta_N$ of Cooper nucleon pair number
$N$ in heated nuclei. The main uncertainty of $E_N$ is caused by the lack of
the experimental data on function $\delta_N=f(U)$, the secondary -- by
uncertainty of  level density of single-particle levels $g$ in model
\cite{Strut}. So, in three different, model dependent approximations of
level density in large set of nuclei of different type \cite{Prep196}
and \cite{PEPAN-2006}, the
threshold $E_2$ of five-quasi-particle excitations  differs by a factor of
1.5-2. 

In practice, for estimation of $K_{vibr}$ from the data \cite{U239} we used the second
variant of notions of correlation function of Cooper pair in heated nucleus
\cite{Prep196}.
 In calculation were used the values $\delta$=0.83 MeV, $g$=14 MeV$^{-1}$.
The best value of breaking threshold for the first Cooper pair was found to be
equal to $E_1=0.55$ MeV when using the assumption on independence of $K_{vibr}$
on nuclear excitation energy. At low excitation energy  this assumption is, most
probably, unreal (see Fig.8).

Parameter $K_{coll}-1$  determined from comparison between the calculated
in this way density of three-quasiparticle excitations (J=1/2, 3/2) and its
most probable experimental value is shown in Fig. 8 together with calculated
value of $\delta_{1}$.

Significant correlation of this coefficient with $\delta_{1}$ from
\cite{PEPAN-2006} and
from the second variant of the analysis \cite{Prep196} is observed
in the excitation energy interval below approximately 1.2 MeV.
At higher excitation energy, decrease in degree of correlation can be related to both considerable contribution of
five-quasiparticle excitations in the function $\rho_{qp}(U,J,\pi)$ and less
than it is accepted in \cite{Prep196,PEPAN-2006}
rate of function $\delta_{1}$ at $U>1.2$ MeV.

The data presented permit one to make the following conclusions:

1. The study of the $^{237}$U nucleus excited in reaction $(\overline{n},\gamma)$ by
neutrons with energies 0.005-0.12, 2 and  24 keV allow us to observe the same
properties as those revealed for $\sim$ 40 nuclei from the mass region
$40 \leq A \leq 200$. There are: step-wise structure in level density
and local enhancement of radiative strength functions of the primary
gamma-transitions to corresponding levels.

2. In the excitation energy region about 0.7-1.5 MeV occurs abrupt change in
level structures. This appears itself in considerable enhancement of  the
$k(M1)/k(E1)$ values and in strong difference of their distribution from normal
distribution of random amplitudes of gamma-transitions.

3. Experimental ratios $k(M1)/k(E1)$ can be used for obtaining more unique
values of strength functions of E1- and M1-transitions and data on relations
between density of levels with different parity in the frameworks of
methods \cite{Meth1,PEPAN-2005}.

4. The majority of the primary gamma-transitions observed in reaction
$(\overline{n},\gamma)$ correspond, probably, to excitation of levels with
large and, maybe, weakly fragmented phonon components of wave functions.

\section{Conclusion}\hspace*{16pt}

Analysis of the available experimental data on the primary gamma-transitions
from reaction $(\overline{n},\gamma)$ in compound nucleus $^{237}$U has
demonstrated step-wise structure in level density and increasing radiative
strength functions of transitions to levels lying in the region of this
structure, at least, for the primary dipole transitions.
I.e., it confirmed main conclusions of \cite{Meth1,PEPAN-2005}.
It showed also a necessity to reveal and remove systematic errors of
experiment in alternative methods for determining only level density and
simultaneously -- all the parameters of the cascade gamma-decay.
Very important for this are both correct accounting for effect of level structures on
probability of emission of evaporated nucleons and cascade gamma-quanta in
investigation of nuclear reactions at beams of accelerators and considerable
reduction of systematic errors of experiment.

Abrupt increase in dispersion (decrease in the $\nu$ parameter) of random
deviations of intensities from the mean (expected according to \cite{PT})
values allows one to suppose the presence in their structure of considerable
components of weakly fragmented nuclear states being more complicated than the
three-quasiparticle states. Approximation of the obtained level density by
Strutinsky model confirms considerable ($\approx$ 10 times) increase in level
density due to excitations of mainly vibration type \cite{PEPAN-2006}.
Comparison of the data presented in figs. 6, 7 with those obtained from
analysis of two-step gamma-cascades permits one to make preliminary
conclusion that the sharp change in structure of decaying neutron resonances,
at least, in their energy interval $\approx$ 24 keV is not observed.
And there are no reasons to expect serious change in determined according
to \cite{Meth1,PEPAN-2005} level density and shape of energy dependence
of radiative strength functions for the primary gamma-transitios from
resonance to resonance. Further reduction of systematic errors of these
nuclear parameters determined from the two-step gamma-cascade
intensities undoubtedly requires reliable estimation of functions
$k(E_\gamma,E_{ex})$ for all energy interval of levels excited at neutron capture.

The analysis  performed in this work and its results point to necessity of
experimental determination of $\rho$ and $k$ by the method \cite{PEPAN-2005}
in all practically important for nuclear energetics nuclei from the
region of actinides.

\end{document}